\documentclass[onecolumn,preprintnumbers,amsmath,amssymb,superscriptaddress]{revtex4}

\usepackage{graphicx}
\usepackage{dcolumn,amssymb,amsmath,amsfonts}
\usepackage{bm}
\usepackage{color}

\begin{document}

\title{The Olami-Feder-Christensen model on a small-world topology}
\author{F. Caruso }
\address{Scuola Superiore di Catania, Via S. Paolo 73, 95123 Catania}
\author{V. Latora , A. Rapisarda }
\address{Dipartimento di Fisica e Astronomia, Universit\`a di Catania, and
INFN sezione di Catania, Via S. Sofia 64, 95123 Catania, Italy}
\author{B. Tadi\'c}
\address{Department  for Theoretical Physics, Jo\v{z}ef Stefan
Institute, P.O. Box 3000; SI-1001 Ljubljana, Slovenia}

\begin{abstract}
We study the effects of the topology on the
Olami-Feder-Christensen (OFC) model, an earthquake model of
self-organized criticality. In particular, we consider a 2D square
lattice and a random rewiring procedure with a parameter $0<p<1$
that allows to tune the interaction graph, in a continuous way,
from the initial local connectivity to a random graph. The main
result is that the OFC model on a small-world topology exhibits
self-organized criticality deep within the non-conservative
regime, contrary to what happens in the nearest-neighbors model.
The probability distribution for avalanche size obeys finite size
scaling, with universal critical exponents in a wide range of
values of the rewiring probability $p$. The pdf's cutoff can be
fitted by a stretched exponential function with the stretching
exponent approaching unity within the small-world region.
\end{abstract}

\maketitle

\section{Introduction}

In order to understand the earthquakes underlying physical
mechanisms, the study and the modelling of seismic space-time
distribution is a fundamental step. In the last years, inspired by
statistical regularities such as the Gutenberg-Richter and the
Omori laws and by a desire to quantify the limits of earthquakes
predictability, a wealth of mechanisms and models have been
proposed. In particular, as a possible explanation for the
widespread occurrence of long-range correlations in space and
time, many authors has modelled the seismogenic crust as a
self-organized complex system, that spontaneously organizes into a
dynamical critical state (SOC) \cite{BTW,bak_book,jen_book}.
Actually the realistic applicability of this kind of models to
describe realistic earthquakes dynamics is still debated
\cite{yang,mega}. A model which has played an important role in
the context of SOC in nonconservative systems is the
Olami-Feder-Christensen (OFC) model of earthquakes~\cite{Olami}.
The most recent numerical investigations have shown that the OFC
model on a square lattice with open boundary conditions displays a
power law distribution of avalanche sizes \cite{lisepac1} with a
universal exponent $\tau \simeq 1.8$, which is independent of the
dissipation parameter; instead in the case with periodic boundary
conditions no power law was found.

Therefore the boundary conditions and in general the underlying
topology play a fundamental role for the criticality; for example,
in the case with open bc, the boundary sites update themselves at
a different frequency from the bulk sites and this inhomogeneity,
together with a divergency length scale in the thermodynamic
limit, induces partial synchronization of the elements of the
systems building up long range spatial correlations and thereby
creating a critical state. However, in this last case, the
distribution there is no finite size scaling. On the other hand,
the OFC model on a quenched random graph \cite{Lise} is
subcritical and shows no power law distributions.
\newline
In this context the purpose of our work is to study the effects of
the graph topology on the criticality of the non-conservative OFC
model. Particularly we consider a small-world graph, following the
method by Watts and Strogatz, and we analyze some SOC properties.
Then we investigate the effects of an increasing number of
long-range connections on the criticality of the model, by varying
the rewiring probability in the range $0<p<1$ . In this case, we
find that for a particular region of values of the structural
parameter $p$ within the small-world regime, the probability
distribution for avalanche size obeys finite size scaling with
universal critical exponents (within numerical error bars).
Moreover, we fit the pdf's cutoffs by a stretched exponential
function and we note that the stretching exponent approaches
$\sigma \to 1$ on the small-world threshold.

\section{The OFC model on a small-world topology}
\label{OFCmodel}

The Olami-Feder-Christensen (OFC) model \cite{Olami} is defined on
a discrete system of $N$ sites on a square lattice, each carrying
a seismogenic force represented by a real variable $F_i$, which
initially takes a random value in the interval $(0,F_{th})$. All
the forces are increased simultaneously and uniformly (mimicking a
uniform tectonic loading), until one of them reaches the threshold
value $F_{th}$ and becomes unstable $(F_i \geq F_{th})$. The
driving is then stopped and an ''earthquake'' (or avalanche)
starts:
\begin{equation}
 \label{av_dyn}
     F_i \geq F_{th}  \Rightarrow \left\{ \begin{array}{l}
      F_i \rightarrow 0 \\
      F_{nn} \rightarrow F_{nn} + \alpha F_i
\end{array} \right.
\end{equation} where ''nn'' denotes the set of nearest-neighbor sites of $i$.
\par The dissipation level of the dynamics is controlled by the
parameter $\alpha$ and, in the case of a graph with fixed
connectivity $q$, it takes values between $0$ and $1/q$ (for the
conservative case $\alpha=1/q$). As regards the model dynamics,
all sites that are above threshold at a given time step in the
avalanche relax simultaneously according to (\ref{av_dyn}) and
then, by a chain reaction, they can create new unstable sites
until there are no more unstable sites in the system ($F_i <
F_{th}$, $\forall i$): this one is an earthquake. At the end of
this event, the uniform growth then starts again and other
earthquakes will happen. The number of topplings during an
earthquake defines its size, $s$, and we will be interested in the
probability distribution $P_N (s)$. In our approach, we point out
that the boundary conditions are "open", i.e. we impose $F=0$ on
the boundary sites.
\newline
Let us point out again that recent numerical investigations have
shown that the non-conservative OFC model on a square lattice
(with "open" boundary conditions) displays scaling behavior, up to
lattice sizes presently accessible by computer
simulations~\cite{lisepac1,lisepac2}. The avalanche size
distribution is described by a power law, characterized by a
universal exponent $\tau \simeq 1.8$, independent of the
dissipation parameter.  This distribution does not display finite
size scaling, however.
\newline
In the OFC model criticality has been ascribed to a mechanism of
partial synchronization~\cite{middleton}. In fact the system has a
tendency to order into a periodic
state~\cite{middleton,socolar,grass2} but is frustrated by the
presence of inhomogeneities such as the boundaries. In particular
these inhomogeneities induce partial synchronization of the
elements of the system building up long range spatial correlations
and a critical state is obtained: therefore the mechanism of
synchronization requires an underlying spatial structure. Indeed,
it is known that the OFC model on a quenched random graph displays
criticality even in the nonconservative regime but introducing
some inhomogeneities \cite{Lise}. In a random graph in which all
the sites have exactly the same number of nearest neighbors $q$
(both for $q=4$ and $q=6$) the dynamics of the NON-CONSERVATIVE
OFC model organizes into a subcritical state. In order to observe
scaling in the avalanche distribution, one has to introduce some
inhomogeneities. It has been found that, for the OFC model on a
(quenched) random graph, it enough to consider just two sites in
the system with coordination $q-1$~\cite{Lise}. In fact, when
either of these sites topple according to rule (\ref{av_dyn}), an
extra amount $\alpha F_i$ is simply lost by the system and a
critical behavior appears.
\newline
Our work consists in the study of the non-conservative OFC model
on a small-world topology and it's motivated by two main reasons.
First of all we expect that the inclusion of some inhomogeneities
in the sites degree is not the unique way to obtain SOC. In fact,
as we are going to show, an alternative way is to keep fixed the
sites degree and to change the topology of the underlying network,
for instance by considering a small world graph, obtained by
randomizing a fraction $p$ of the links of the regular nearest
neighbor lattice. The method to construct networks with
intermediate properties between regular and random graphs proposed
by Watts and Strogatz \cite{watts,wattsbook} is perfectly suited
to our purpose since it allows us to consider networks that are in
between the two most widely investigated cases, namely the nearest
neighbors lattice and the random graph. The second reason has to
do with modelling the real space-time seismicity. A small-world
topology is expected to be a more accurate description of a real
systems according to the most recent geophysical observations that
indicate that earthquakes correlation might extend to the long
range in both time and space. In fact, if a main fracture episode
occurs, it may induce slow strain redistribution through the earth
crust, thus triggering long-range as well as short-range seismic
effects \cite{kagan,hill,cresce,parsons,palatella}. The presence
of a certain percentage of long-range connections in the network
takes into account the possibility that an earthquake can trigger
earthquakes not only locally but also at distant regions. Actually
by following the method proposed by Watts and Strogatz we start
with a two-dimensional square lattice in which each site is
connected to its $4$ nearest neighbors and then the links of the
lattice are rewired at random with a probability $p$. The main
differences with respect to the original model is that for any
value of $p$ we want to keep fixed the connectivity of each site
and therefore the connections are rewired in couples.

\section{Results}
\begin{figure} [!]
\begin{center}
\includegraphics[width=8.cm,angle=0]{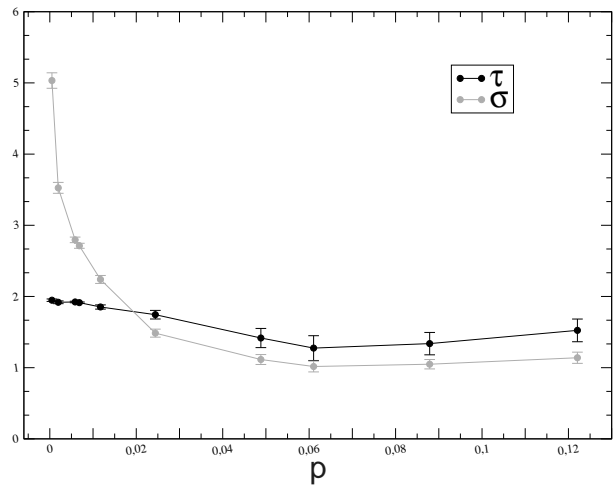}
\caption{\label{tausigma} The two exponents $\tau$ and $\sigma$ as
a function of the rewiring parameter $p$; this results refer to
the case with L=$64$ and in the other cases we get the similar
results.}
\end{center}
\end{figure}

In our simulations we have studied a two-dimensional square
lattice (with "open" boundary conditions) $L \times L$ with three
different sizes: $L=32, 64$ and $128$. The corresponding number of
sites is $N= L^2$. We have considered up to $1E+09$ avalanches to
obtain a good statistics for the avalanche size distribution
$P_N(s)$.
\newline
All the curves can be fitted by a stretched-exponential function:
\begin{equation}
\label{fss} P_N(s)=A s^{-\tau} e^{-(s/\xi)^\sigma}
\end{equation}
where $s$ is the avalanche size, $\xi$ is the characteristic
length and $\tau$ and $\sigma$ are two exponents. We notice that,
on approaching the small-world limit, the size distribution
power-law is practically lost and the stretching becomes more and
more pronounced. This can be better exploited by plotting the
value of the two exponents $\tau$ and $\sigma$ as a function of
$p$ in Fig.\ref{tausigma}. In this figure $\tau$  takes the value
1.8 for $p=0.02$ at the small-world transition. \par Moreover we
observe a sudden change in the behavior of the stretching exponent
$\sigma$ at SW probability. In general one can expect
stretching-exponential in various cases of stochastic processes
where many length scales appear. So, when the system is at the
small-world threshold (and beyond) multiple lengths start playing
a role, because of presence of long-range links.

In order to characterize the critical behavior of the model, a
finite size scaling (FSS) ansatz is used, i.e.
\begin{equation}
\label{fss} P_N(s) \simeq N^{-\beta} f(s/N^D)
\end{equation}
where $f$ is a suitable scaling function and $\beta$ and $D$ are
critical exponents describing the scaling of the distribution
function.
\par
In Fig.\ref{fss} we consider $\alpha=0.21$ and rewiring
probability $p=0.006$ below the small-world transition and we show
the collapse of $P_N(s)$ for three different values of $N$, namely
$N=32^2, 64^2, 128^2$. We find that the distribution $P_N(s)$
satisfies the FSS hypothesis reasonably well, with universal
critical coefficients with small rewiring probability, but, on
approaching the small-world limit ($p=0.02$), the cut-offs are not
exponential, so no FSS is expected. The critical exponent derived
from the fit of Fig.\ref{fss} are $\beta \simeq 3.6$ and $D=2$,
independent of the dissipation parameter $\alpha$. The FSS
hypothesis implies that, for asymptotically large $N$, $P_N(s)
\sim s^{-\tau}$ and the value of the exponent is $\tau= \beta/D
\simeq 1.8$. Because of the numerical errors it is difficult to
assert with certainty that $\tau$ is a novel exponent, different
from the one for the conservative RN model ($\tau=1.5$). However
the power law exponent of the distribution is consistent with the
exponent of the OFC model with open boundary conditions, i.e.
$\tau \simeq 1.8$.
\par
On a rewired topology the system behaves as in the compact square
lattice, but the occurrence of a small amount of long-range links
disseminate the avalanches over the network and the whole
avalanche has the property to scale with the systems size; in fact
the links, randomly long, can be system-size long. A superposition
of all these local events could be one of possible explanations of
this whole instability (avalanche) and then the stretching
exponential cut-offs can be used to fit the size avalanche
distribution.

\begin{figure} [t]
\begin{center}
\includegraphics[width=8.2cm,angle=0]{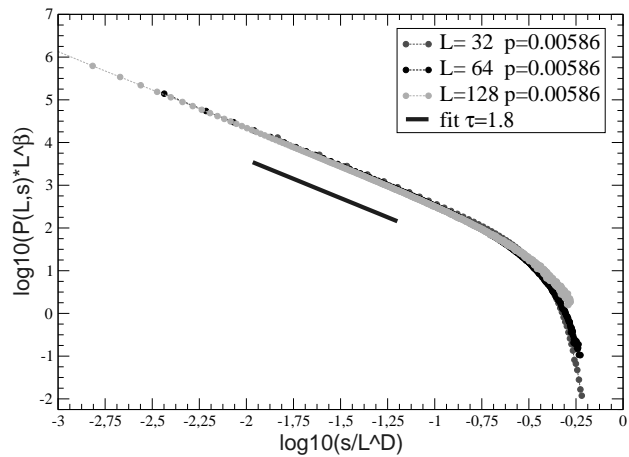}
\caption{\label{fss} Finite size scaling for dissipative OFC model
on a small world topology for three different values of $N$,
namely $N=32^2, 64^2, 128^2$; by fitting the curves, we derive the
following the critical exponents $D=2$ and $\beta \simeq 3.6$.}
\end{center}
\end{figure}

\section{Conclusions}
\label{Conclusions} In this work we have studied how the topology
can play an important role on the dynamics of the OFC model.
Following this idea a rewiring procedure as that introduced by
Watts and Strogatz \cite{watts} has been considered and then we
have found that the model is critical even in the nonconservative
regime. In fact, respect to the NN OFC model, in which the
criticality appears only in the conservative case unless
introducing any inhomogeneities in some sites, a small world graph
has an underlying spatial structure so that partial
synchronization of the elements of the system can still occur,
without being destroyed as in the quenched random graph without
inhomogeneities. Finally we point out that, on approaching the
small-world limit, the power-law is practically lost, the cutoffs
are fitted by a stretched-exponential function and the sudden
change of behavior of the stretching exponent $\sigma$ appears at
SW probability. Moreover we should emphasize that this kind of
study can be easily extended to other topologies, as scale-free
networks, in order to understand better and better the role of the
topology for the earthquakes models.

\end{document}